\documentclass[aps, pra, twocolumns, reprint]{revtex4-2}

\usepackage{amssymb,amsmath,latexsym,amscd,amsfonts, bbm, bm}  
\usepackage{graphicx}  
\usepackage[normalem]{ulem}
\usepackage[usenames,svgnames]{xcolor}
\usepackage[lined,algoruled]{algorithm2e}

\usepackage{hyperref}
\hypersetup{
     colorlinks=true,           
     linkcolor=DarkCyan,          
     citecolor=blue,            
     filecolor=blue,            
     urlcolor=cyan,             
 }

\graphicspath{{Figs}}

\newtheorem{theorem}{Theorem}[section]

\newtheorem{lemma}[theorem]{Lemma}

\newtheorem{problem}[theorem]{Problem}  

\newcommand{\Eq}[1]{Eq.~(\ref{#1})}
\newcommand{\Fig}[1]{Fig.~\ref{#1}}

\newcommand{\Lem}[1]{Lemma~\ref{#1}}

\newcommand{\Sec}[1]{Sec.~\ref{#1}}
\newcommand{\cRef}[1]{Ref.~\cite{#1}}
\newcommand{\cRefs}[1]{Refs.~\cite{#1}}
\newcommand{\cc}[1]{~\cite{#1}}

\newcommand{\qedsymb}{\hfill{\rule{2mm}{2mm}}}

\newcommand{\ignore}[1]{}

\newcommand{\ket}[1]{{ |{#1} \rangle }}  

\newcommand{\av}[1]{{ \langle {#1} \rangle }}

\newcommand{\ketbra}[2]{{ |{#1} \rangle\langle {#2} | }}

\newcommand{\EqDef}{\stackrel{\mathrm{def}}{=}}

\newcommand{\eps}{\epsilon}

\DeclareMathOperator*{\Tr}{Tr}

\newcommand{\mcP}{\mathcal{P}}
\newcommand{\mcF}{\mathcal{F}}

\newcommand{\mcG}{\mathcal{G}}
\newcommand{\mcH}{\mathcal{H}}
\newcommand{\mcN}{\mathcal{N}}
\newcommand{\mcT}{\mathcal{T}}

\newcommand{\us}{{\underline{s}}}
\newcommand{\Bx}{\bm{x}}

\newcommand{\blockBP}{\texttt{blockBP }}

\begin{document}

\title{A blockBP decoder for the surface code} 

\author{Aviad Kaufmann}
\affiliation{Physics Department, Technion – Israel Institute of
Technology, 3200003, Haifa, Israel}
\email{aviadskaufmann@gmail.com}

\author{Itai Arad}
\affiliation{Centre for Quantum Technologies, National University of
Singapore, 117543 Singapore, Singapore}

\date{\today}

\begin{abstract}
  We present a new decoder for the surface code, which combines the
  accuracy of the tensor-network decoders with the efficiency and
  parallelism of the belief-propagation algorithm. Our main idea is
  to replace the expensive tensor-network contraction step in the
  tensor-network decoders with the \blockBP algorithm --- a recent
  approximate contraction algorithm, based on belief propagation.
  Our decoder is therefore a belief-propagation decoder that works
  in the degenerate maximal likelihood decoding framework. Unlike
  conventional tensor-network decoders, our algorithm can run
  efficiently in parallel, and may therefore be suitable for
  real-time decoding. We numerically test our decoder and show that
  for a large range of lattice sizes and noise levels it delivers a
  logical error probability that outperforms the MWPM decoder,
  sometimes by more than an order of magnitude. 
\end{abstract}

\maketitle

\section{Introduction}

The quantum surface code\cc{ref:Kitaev2003-surface} is one of our
best candidate codes for building a fault-tolerant quantum
computer\cc{ref:AB2008-FT}, owing to its high degree of locality,
high threshold and large code distance. It is a well-studied code,
and simple demonstrations of it have already been
performed\cc{ref:ETH2022-surface-FT, ref:Chinese2022-surface-FT,
ref:Google2023-FT-exp}. As the current and near-future quantum
hardware will probably be characterized by a relatively low number
of qubits (few hundreds to few thousands), with a not-so-low noise
levels, it is important to optimize the use of this code as much as
possible. This is a highly non-trivial task. It requires deep
understanding and integration of the many components that take part
in error-correction process, software and hardware alike.

An important ingredient in the fault-tolerance framework is the
decoder. This is a classical algorithm that takes as input the
results of the syndrome measurements and outputs a
description of a corrective action in the form of a Pauli
string. Finding the optimal corrective action is a hard
task\cc{ref:IP2015-hardness}, as one needs to look over an
exponentially large space of possible solutions. In addition, the
decoder has to be highly efficient, as it has to run in time scales
that are short enough for the system not to decohere too much ---
scales of the order of $1\mu\, \text{sec}$ for superconducting
qubits, for example.

Over the years, a plethora of surface codes decoders have been
developed\cc{ref:Orus2023-surfaceD}, using various approaches. In
general, these decoders follow the natural trend in which the
decoder accuracy anti-correlates with its running time. At the
extreme side of this spectrum lie tensor-network (TN) based
decoders\cc{ref:BSMV2014-TN-decoding, ref:TDCBBF2019-TNdec,
ref:CF2021-TNdec, ref:Chubb2021-TNdec, ref:CCR2023-TNdec}.  They 
have a nearly optimal decoding accuracy that comes at a very high
computational cost, which makes them unpractical for real-time
decoding. 

In this paper, we initiate the study of a new TN-based decoder,
which on one hand enjoys high accuracy, but on the other hand, being
a parallel algorithm, it can potentially be used for real-time
decoding. Our idea is to use the recent \blockBP
algorithm\cc{ref:GPA2023-blockBP} as the contraction engine for the
tensor-network decoder.  \blockBP is an algorithm for approximate
contraction of 2D tensor networks, which is based on the
belief-propagation (BP) algorithm\cc{ref:Pearl1982-BP}, in
conjunction with matrix product states techniques. As a BP
algorithm, it can be easily parallelized, making it suitable for
real-time decoding. Indeed, BP algorithms are at the backbone of
some of the most powerful classical codes decoders such as LDPC
codes\cc{ref:SFRU2001-BP-LDPC} and Turbo
codes\cc{ref:BTG1993-BP-Turbo}. They have also been extensively
studied in the context of quantum
decoding\cc{ref:PC2008-BP-vanilla}, in particular in conjunction
with other decoders\cc{ref:DP2010-RG-decoder, ref:DP2014-RG-decoder,
ref:CA2018-belief-matching, ref:HBKFC2023-belief-matching}. Our use
of the BP algorithm is different in that we use it for the task of
tensor-network contraction, which results in a BP algorithm for
degenerate quantum maximal likelihood decoding.

To demonstrate our method, we test it in the prefect measurement
setup (i.e., the code capacity model) using surface codes of various
sizes with a single-qubit depolarizing noise of different levels. We
compare its performance to the original
TN\cc{ref:BSMV2014-TN-decoding} and the Minimal Weight Perfect
Matching (MWPM)\cc{ref:Edmonds1965-MWPM, ref:DKLP2003-FT,
ref:FMMC2012-MWPM, ref:Fowler2015-MWPM} decoders. We believe that
our techniques can be straightforwardly generalized to more
realistic noise models, including circuit-level noise, as well as to
other topological two-dimensional codes such as the color
code\cc{ref:KYP2015-color-code} or the rotated surface
code\cc{ref:BM2007-rotated}. 

The organization of the paper is as follows. In \Sec{sec:background}
we offer a short introduction to stabilizer codes and their decoding
problem, and describe the surface code. In \Sec{sec:blockBP} we
describe our \blockBP decoder. We include also a short description
of the original TN-based decoder, as well as the BP algorithm for TN
contraction. In \Sec{sec:numerics} we present the results of our
numerical simulations, and then offer our conclusions in
\Sec{sec:conclusions}.

\section{Theoretical Background}
\label{sec:background}

\subsection{Maximal likelihood and  degenerate maximal likelihood
  decoding in stabilizer codes}

To present the formalism of stabilizer
codes\cc{ref:Gottesman1997-stabilizers} and the decoding problem, we
largely follow the notation of \cRef{ref:BSMV2014-TN-decoding}.

Let $\mcP_n$ denote the Pauli group over $n$ qubits. Every member of
this group is an operator $f = cf_1\otimes f_2\otimes\ldots\otimes
f_n$, where $f_i \in \{I, X, Y, Z\}$ are single-qubit Pauli
operators and $c\in\{\pm 1, \pm i\}$. A stabilizer code is defined
by a \emph{stabilizer subgroup} $\mcG\triangleleft \mcP_n$ that is
(i) abelian and (ii) does not contain the $-I$ element.  The
condition $-I\notin \mcG$ implies that the $c$ prefactors in the
expansion of every $g\in \mcG$ must be $\pm 1$, and therefore all
$g\in \mcG$ have $\pm 1$ eigenvalues.

In what follows, it will be beneficial to discuss $\mcG$ in terms of
its generators, which are also referred to as ``check operators'' in
the context of stabilizer codes. The generators are a \emph{minimal}
set of elements $g_1, g_2, \ldots, g_m\in\mcG$ such that $\av{g_1,
\ldots, g_m} = \mcG$. As the different $g_i$ commute among
themselves, they share a common eigenbasis.  The stabilizer subspace
$\mcH_0$ is defined by $\mcH_0 \EqDef \{\ket{\psi} : g_i\ket{\psi} =
\ket{\psi} , \quad \forall \ i=1, \ldots m\}$, i.e., it is the span of
eigenvectors of $g_i$ with common eigenvalue $1$. It can be shown
that $\dim \mcH_0 = 2^k$, where $k\EqDef n-m$ is the number of
logical qubits that can be encoded in
$\mcH_0$\cc{ref:DSS1998-Stabilizer}.

In the QECC protocol, we start with an encoded state $\rho\in
\mcH_0$ that experiences a noise described by a noise channel
$\mcN$. In this work, we assume $\mcN$ is a stochastic Pauli channel
\begin{align}
\label{eq:noise}
  \mcN(\rho) \EqDef \sum_{f\in \mcP_n} \pi(f) \cdot f \rho f^\dagger 
\end{align}
with known probabilities $\pi(f)$. Our task is to revert the affect
of the noise and bring the system back to the encoded state $\rho$.

We start by simplifying \Eq{eq:noise} using an important
observation. As $\rho\in \mcH_0$, it is easy to see that $f_1\rho
f_1^{-1} = f_2\rho f_2^{-1}$ for any $f_1, f_2$ for which
$f_1^{-1}f_2\in \mcG$. In other words, there are many different
errors/operators that have the \emph{same} action over the code
space. This property is known as \emph{degeneracy}, and it is a
genuine property of quantum codes that does not exist in classical
codes. Mathematically, we can rephrase it by saying that $f_1,f_2$
have the same action on the code space iff they belong to the same
\emph{coset} of $\mcG$ in $\mcP_n$, i.e., $f_1\mcG =f_2\mcG$.
Therefore, letting $\{f_\alpha\}$ denote the different coset
representatives, \Eq{eq:noise} becomes
\begin{align}
\label{eq:noise-cosets}
  \mcN(\rho) = \sum_\alpha 
    \pi(f_\alpha\mcG) \cdot f_\alpha\rho f_\alpha^\dagger .
\end{align}
It can be interpreted as saying that the noise channel takes $\rho$
to one of the ``coset states'' $f_\alpha\rho f_\alpha^\dagger$ with
probability $\pi(f_\alpha\mcG) \EqDef \sum_{g\in \mcG} \pi(f_\alpha
g)$. If we could perform a measurement that would tell us at which
coset the state is, we could simply apply $f_\alpha^\dagger$ (or any
other state in that coset) to take us back to $\rho$. Unfortunately,
this is impossible, as the $f_\alpha\rho f^\dagger_\alpha$ states
are not necessarily orthogonal to each other.  Instead, we perform a
\emph{syndrome measurement}, in which we measure the eigenspaces of
the $m$ generators $g_i$, with each measurement giving us either
$+1$ or $-1$. Since all the generators commute with each other, they
define a decomposition of $\mcH$ into a direct sum $\mcH =
\bigoplus_\us \mcH_\us$ of common eigenspaces. The indices $\us$ are
binary strings $\us = (s_1, \ldots, s_m)$, called \emph{syndromes},
which define the eigenvalues of the generators on $\mcH_\us$:
$\ket{\psi}\in \mcH_\us$ iff $g_i\ket{\psi} = (-1)^{s_i}\ket{\psi}$.
In that notation, the stabilizer subspace $\mcH_0$ corresponds to
the syndrome $\us=(0,\ldots, 0)$.

After measuring the stabilizers, we obtain the syndrome $\us=(s_1,
\ldots, s_m)$ of the subspace $\mcH_\us$ into which the system 
collapsed. As the Pauli operators either commute or anti-commute, 
$f\mcH_0f^\dagger = \mcH_\us$ iff $f_\alpha g_i =
(-1)^{s_i}g_if_\alpha$ for $i=1,\ldots,m$. Therefore, every $f\in
\mcP_n$ maps $\mcH_0$ to a well-defined $\mcH_\us$. This defines yet
another partition of $\mcP_n$ into $2^m$ sets of Pauli operators
that take $\mcH_0$ to the same $\mcH_\us$. Mathematically, it is the
partition of $\mcP_n$ into the cosets of the \emph{centralizer}
group of $\mcG$, which is defined by elements in $\mcP_n$ that
commute with all elements of $\mcG$:
\begin{align}
  C(\mcG) \EqDef \{ f\in \mcP_n : fg=gf \ \forall g\in \mcG\}.
\end{align}
Indeed, given two Pauli operators $f_a, f_b$, then $f_a g_i =
(-1)^{s_i} g_i f_a$ and $f_b g_i = (-1)^{s_i} g_i f_b$ for all $i$
iff $f_a f_b^{-1}$ commutes with all $g_i$, which happens iff $f_a
f_b^{-1} \in C(\mcG)$. To summarize, there are $2^m$ cosets of
$C(\mcG)$ in $\mcP_n$, which correspond to the sets of Pauli operators
that take $\mcH_0$ to any of $2^m$ subspaces $\mcH_\us$. As $\mcG
\triangleleft C(\mcG)$, every coset of $C(\mcG)$ can be broken into 
$|C(\mcG)\backslash \mcG|=2^{2k}$ cosets  of $\mcG$. 

Let $L_0, L_1, L_2\ldots$ denote the representatives of the cosets
of $\mcG$ in $C(\mcG$) so that
\begin{align*}
  C(\mcG) = L_0 \mcG \, \cup\, L_1\mcG \, \cup\, \ldots 
\end{align*}
These operators correspond to logical gates on the encoded space. We
use them to write every coset of $C(\mcG)$ that is associated with a
syndrome $\us$ as a disjoint union of $\mcG$ cosets:
\begin{align*}
  f_\us C(\mcG) = f_\us L_1\mcG \, \cup\, f_\us L_2\mcG\, \cup
  \ldots .
\end{align*}
The post measurement state after measuring a syndrome $\us$ is
therefore
\begin{align}
  \rho_\us \propto \sum_\beta \pi(f_\us L_\beta \mcG)\cdot
    f_\us L_\beta \rho L^\dagger_\beta f^\dagger_\us  .
\end{align}
We can now summarize the stabilizer code decoding problem as follows:
\begin{problem}[decoding a stabilizer code]
  Given the result of a syndrome measurement $\us=(s_1, \ldots,
  s_m)$, find a corrective Pauli operator $f$ that will take
  $\rho_\us$ back to $\rho$ with high probability.
\end{problem}

From the above discussion it is clear that the best guess for a
corrective action given a syndrome $\us$ would be to find the coset
$f_\us L_\beta \mcG$ with the maximal probability $\pi(f_\us L_\beta
\mcG)$ and then take any $f\in f_\us L_\beta \mcG$. This type of
decoding is called \emph{Degenerate Quantum Maximal Likelihood
Decoding (DQMLD)}. It is the approach taken by tensor-network based
decoders\cc{ref:BSMV2014-TN-decoding, ref:TDCBBF2019-TNdec,
ref:CF2021-TNdec, ref:Chubb2021-TNdec, ref:CCR2023-TNdec} (see
\Sec{sec:surface-to-TN} for more details), as well as (to a lesser
extent) by RG-based decoders\cc{ref:DP2010-RG-decoder,
ref:DP2014-RG-decoder}. As was shown in \cRef{ref:IP2015-hardness},
it is a \#P-complete problem, and therefore in the worst case we can
only try to solve it approximately. 

The reason why DQMLD is harder than classical decoding is that in
order to calculate the coset probability $\pi(f_\us L_\beta \mcG)$,
we need to sum over an exponential number of probabilities of the
Pauli operators in it. A simpler, yet less optimal approach would be
to ignore the degeneracy of the code and look for the noise operator
$f\in \mcP_n$ that maximizes $\pi(f)$. This type of decoding is
called \emph{Quantum Maximal Likelihood Decoding (QMLD)}, and it is
the approach taken by most ``practical'' decoders such as the
Minimal Weight Perfect Matching (MWPM)
decoder\cc{ref:Edmonds1965-MWPM, ref:DKLP2003-FT, ref:FMMC2012-MWPM,
ref:Fowler2015-MWPM}, the Union Find decoder\cc{ref:DN2021-UNdec}
and variants of BP-based decoders\cc{ref:PC2008-BP-vanilla,
ref:CA2018-belief-matching, ref:HBKFC2023-belief-matching,
ref:PK2021-BP-OSD, ref:RWBC2020-BP-OSD}.  While it is an easier
problem than DQMLD, it is still a hard problem, which has been shown
to be NP-complete\cc{ref:HL2011-NP-hardness, ref:KL2020-NP-hardness,
ref:KL2012-NP-hardness}, as its classical
counterpart\cc{ref:BMT1978-classicNP}.

In many practical scenarios, in particular when the noise level is
very low, QMLD would yield very good results, approaching the
performance of DQMLD decoding\cc{ref:IP2015-hardness}. Nevertheless,
empirical tests show that often also in low error regime, DQMLD
encoders such as the TN-based decoders provide logical error rates
that can be orders of magnitude lower than QMLD based decoders such
as MWPM (see, for example the results in \Sec{sec:numerics}).

\subsection{The surface code}
\label{sec:surface}

\begin{figure}
  \begin{center}
    \includegraphics[scale=1]{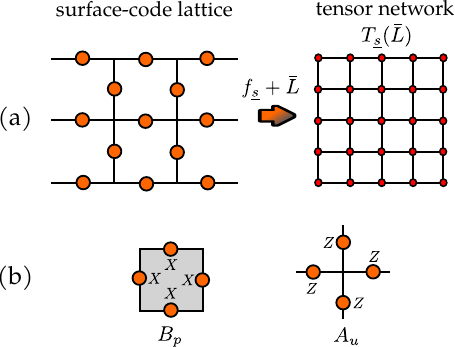}
  \end{center}
  \caption{(a) The lattice structure of a $d=3$ surface code. The
  qubits sit on the edges of the square lattice with $d^2$
  horizontal edges and $(d-1)^2$ vertical edges. Given an Pauli
  operator $f_\us$ that corresponds to the syndrome $\us$ and a
  logical operator $\bar{L}\in \{\bar{I}, \bar{X}, \bar{Y},
  \bar{Z}\}$, the lattice is mapped to a $(2d-1)\times(2d-1)$ square
  tensor network $T_\us(\bar{L})$, whose contraction value gives the
  coset probability $\pi(f_\us \bar{L}\mcG)$. See
  \Sec{sec:surface-to-TN}. (b) The check operators (stabilizers) of
  the surface code: Plaquette operators $B_p$ are the product of
  Pauli $X$ operators sitting on the edges surrounding a plaquette
  $p$.  Site operators $A_u$ are the product of Pauli $Z$ operators
  on the qubits that sit on the edges adjacent to a site $u$.}
  \label{fig:surface-code} 
\end{figure}

The surface code\cc{ref:Kitaev2003-surface} is a stabilizer code
that can be defined on any two-dimensional lattice that is described
by a graph $G=(V,E)$. For simplicity, we shall limit our discussion
to an open square lattice of size $d\times d$, but much of the
discussion below can be generalized to other 2D lattices. Following
the convention of \cRef{ref:BSMV2014-TN-decoding}, our lattice is
made of $d^2$ horizontal edges and $(d-1)^2$ vertical edges, as
shown in \Fig{fig:surface-code}(a) for the case of $d=3$. The qubits
sit on the edges so that in total there are $n=d^2 + (d-1)^2$
qubits. There are two types of check operators: Z check operators,
denoted by $A_u$ and X check operators that are denoted by $B_p$,
see \Fig{fig:surface-code}(b). The $A_u$ check operators are defined
for every site $u\in V$ as the product of Pauli $Z$ on its adjacent
edges. $B_p$ is defined for every plaquette $p$ as the product of
Pauli $X$ operators on its boundary:
\begin{align}
\label{def:AB}
  A_u &\EqDef \prod_{u\in e} Z_e, &
  B_p &\EqDef \prod_{e\in p} X_e .
\end{align}
There are $d(d-1)$ check operators of each type. The $A_u$ check
operators trivially commute within themselves, and so do the $B_p$
operators. It is also easy to see that the two types of operators
commute with each other as every pair either intersect on an even
number of edges or does not intersect at all.

With the $m=2d(d-1)$ independent check operators defined above over
$n=d^2 + (d-1)^2$ qubits, we have $k=n-m=1$, which corresponds code
space $\mcH_0$ that encodes a single logical qubit. The logical
operators $\bar{L}_i$ in this case correspond to logical Paulis on
the encoded qubit and are denoted by $\{ \bar{I}, \bar{X}, \bar{Y},
\bar{Z}\}$. The logical $\bar{X}$ [$\bar{Z}$] operators 
can be written as Pauli $X$  [$Z$] operators over a row [column] of
horizontal edges.

\section{The blockBP decoder for the surface code}
\label{sec:blockBP}

The blockBP decoder is based on the tensor-network
decoder\cc{ref:BSMV2014-TN-decoding}, which we will first review
briefly here. Throughout this section we assume basic knowledge of
tensor networks and matrix product states in particular. Good
reviews of these subjects can be found, for example, in
\cRefs{ref:Orus2014-practical-TN, ref:Schollwock2011-MPS}

\subsection{The Tensor Network based DQMLD decoder}
\label{sec:surface-to-TN}

In \cRef{ref:BSMV2014-TN-decoding} it was shown that the DQMLD
problem for the $d\times d$ surface code can be reduced to the
problem of contracting a $(2d-1)\times (2d-1)$ square tensor-network
with a bond dimension of $D=2$. Specifically, given a syndrome
measurement result $\us$, one can efficiently construct 4 such
square tensor networks, which we denote by $T_\us(\bar{I}),
T_\us(\bar{X}), T_\us(\bar{Y}), T_\us(\bar{Z})$. The contraction of
these networks is equal to the coset probabilities $\pi(f_\us \mcG),
\pi(f_\us \bar{X}\mcG), \pi(f_\us \bar{Y}\mcG), \pi(f_\us
\bar{Z}\mcG)$, from which we can pick the coset with the highest
probability. This reduces the DQMLD problem to the problem of
contracting a 2D TN.  Unfortunately, contracting a 2D TN is a
\#P-hard problem\cc{ref:SWVC2007-2D-contraction}, and therefore we
can generally only solve it approximately. 

In \cRef{ref:BSMV2014-TN-decoding}, the authors used the boundary
MPS method (bMPS)\cc{ref:VC2004-bMPS, ref:VMC2008-PEPS} to
approximate the TN contraction. In this method, the 2D TN is
contracted column by column from left to right using the matrix
product state (MPS) machinery\cc{ref:Schollwock2011-MPS}. At step
$i$, the contraction of the TN up to column $i$ is approximated by a
MPS with a finite bond dimension $\chi$. Next, viewing the $i+1$
column as a matrix product operator (MPO), it is contracted with the
MPS, creating a new MPS with a bond dimension of $2\chi$. To keep
the bond dimension finite, the new MPS is compressed by truncating
back the bond dimension to $\chi$ using well-established MPS
methods\cc{ref:Schollwock2011-MPS}.

Being a DQMLD decoder, the bMPS decoder gives the best decoding
results of all efficient (i.e., with polynomial running time)
decoding algorithms, and has the highest thresholds for a large
range of parameters\cc{ref:Orus2023-surfaceD}. However, for a fixed
truncation bond $\chi$, the computational cost of the algorithm is
$O(n\chi^3)$.  For the decoder to succeed, the error in the TN
approximation should be smaller than the typical difference between
the coset probabilities. This forces $\chi$ to increase as $d$
increases, in particular when we get closer to the noise threshold.
This, together with a large overhead in the $O(\cdot)$ notation,
makes the bMPS a very slow decoder that is challenging to implement
in real-time scenarios. Indeed, in \cRef{ref:Google2023-FT-exp}, as
well as in the numerical experiments we have conducted in this
paper, the bMPS decoder was many orders of magnitude slower than the
MWPM decoder.

\subsection{BP decoders for quantum codes}

Our main idea is to replace the slow bMPS contraction with the
blockBP algorithm\cc{ref:GPA2023-blockBP}. It is a Belief
Propagation (BP)\cc{ref:Pearl1982-BP, ref:MM2009-BPbook} algorithm
for an approximate contraction of 2D tensor networks. BP is a
message passing meta-algorithm, which was originally
designed for approximating the marginals of multivariate
probability distributions defined over complex networks.
It is exact on trees and often gives surprisingly good results also
for loopy networks. Nowadays, it is used in diverse areas such as
Bayesian inference\cc{ref:Pearl1982-BP}, statistical
physics\cc{ref:MPVM1987-SpinGlass, ref:YFDM2011-BP-Ising,
ref:KNZ2014-Percolation}, and combinatorial
optimizations\cc{ref:MPZ2002-BP-SAT, ref:MM2009-BPbook}, to name a
few. It is also the backbone of powerful decoding algorithms for
classical LDPC and Turbo codes\cc{ref:SFRU2001-BP-LDPC,
ref:BTG1993-BP-Turbo}, where it often referred to as the
\emph{sum-product algorithm}.

We note that BP is also used in quantum decoders.  As in the
classical LDPC case, BP can be used to solve the QMLD problem by
trying to find the most probable error. However, applying this
approach to the surface code (and other topological codes) results
in very poor performance.  As argued by
\cRef{ref:PC2008-BP-vanilla}, the failure can be attributed to the
presence of degeneracy and many small loops in the network over
which BP runs. However, we note that more recently there have been
several successful approaches of using BP as a quantum decoder by
combining it with other decoders\cc{ref:DP2010-RG-decoder,
ref:DP2014-RG-decoder, ref:CA2018-belief-matching,
ref:HBKFC2023-belief-matching} or with some post
processing\cc{ref:PK2021-BP-OSD, ref:RWBC2020-BP-OSD}.

\subsection{Using BP for tensor network contraction}

We use BP as an approximate TN contraction method to estimate the
different probabilities in the DQMLD framework. The idea of using BP
for approximate TN contraction was first introduced in
\cRef{ref:AA2021-TNBP}, and subsequently used in various places for
achieving a fast TN contraction\cc{ref:SS2022-BPupdate,
ref:TJG2024-IBMsim}, or for finding a suitable local TN
gauge\cc{ref:TFSMS2024-TNBP-IBM, ref:TM2023-BPgauge}. In
\cRef{ref:GPA2023-blockBP} this approach was further strengthened
for planar TN using the MPS machinery together with coarse
graining in an algorithm called \blockBP. We briefly sketch this
algorithm here. For more details, the reader is referred to
\cRefs{ref:MM2009-BPbook, ref:AA2021-TNBP, ref:GPA2023-blockBP}.

\begin{figure}
  \begin{center}
    \includegraphics[scale=1]{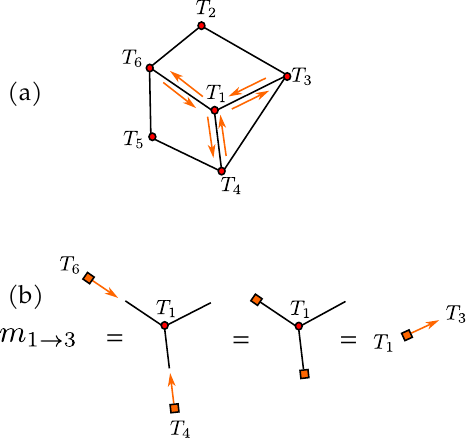}
  \end{center}
  \caption{Sketch of the BP algorithm for tensor networks. (a) a
  tensor network with 5 tensors. Each neighboring tensors exchange
  messages between them. (b) The outgoing message from $T_1$ to
  $T_3$ is obtained by contracting $T_1$ with all the incoming
  messages to $T_1$ from the previous round except for the incoming
  message from $T_3$.} \label{fig:BP} 
\end{figure}

We first describe the ``vanilla BP'' algorithm of
\cRef{ref:AA2021-TNBP} and then explain how it can be improved by
using the block structure of \cRef{ref:GPA2023-blockBP}. Consider
then a TN $\mcT$ that is described by a graph $G=(V,E)$ with a
tensor $T_v$ on every vertex $v\in V$. The incident edges to vertex
$v$ correspond to the different indices of $T_v$. An edge connecting
$u$ to $v$ indicates a contraction of the tensors $T_u, T_v$ along
the two indices of that edge, see \Fig{fig:BP}(a). Our goal is to
approximate the contraction of $\mcT$, which we denote by
$P=\Tr(\mcT)$. As explained in \cRef{ref:AA2021-TNBP} (see also
\cRef{ref:Robeva2017-TN-Duality}), this contraction can be
approximated using a simple BP algorithm. The algorithm is sketched
in \Fig{fig:BP}. It contains several rounds of message passing
between neighboring nodes on the graph. For every edge $e=(u,v)$ on
the graph there will be two messages in the opposite directions:
$m_{u\to v}(x_e)$ and $m_{v\to u}(x_e)$, where $x_e$ is a discrete
variable that runs over the index of the common leg $e$ of the
tensors $T_u,T_v$. As such, $m_{v\to u}(x_e)$ can be viewed as a
vector, or as a tensor with a single leg. At each round, a node $v$
uses the incoming messages of the previous round (sent to him by its
neighboring nodes) to calculate an outgoing message for each one of
its neighbors. Formally, let $N_v$ denote all the neighboring nodes
of $v$ on the graph, and let $m_{u\to v}^{(t)}$ denote the message
from node $u$ to node $v$ at round $t$. Then the message update
rule, which is illustrated in \Fig{fig:BP}(b), is given by
\begin{align}
\label{eq:BP-update}
  m^{(t+1)}_{v\to u} = \Tr(T_v\!\!\!\!\!\!\!\!
    \prod_{u'\in N_v\setminus\{u\}}\!\!\!\!\!
    m^{(t)}_{u'\to v}) 
\end{align}
where $\Tr(\cdot)$ stands for contraction over internal edges in the
TN.

Once the messages have converged to a fixed point, or once the
number of rounds exceeds some prescribed limit, the resulting
messages are used to approximate $P$. One uses the converged
messages to calculate the \emph{Bethe
free-energy}\cc{ref:MM2009-BPbook, ref:AA2021-TNBP} functional
$\mcF_{Bethe}(\mcT)$ of the tensor network from the converged
messages and then approximates $\Tr(\mcT) \simeq
e^{-\mcF_{Bethe}(\mcT)}$. However, a direct use of the Bethe
free-energy formula will be highly inefficient, in particular in the
blockBP framework that we describe later. Instead, the following
lemma provides an equivalent formula, which can easily used in TN
framework:
\begin{lemma}
\label{lem:Bethe}
  Let $\mcT$ be a TN defined on a graph $G=(V,E)$, and let $m_{u\to
  v}(x_e)$ be the set of converged BP messages. Let $\mcF_{Bethe}(\mcT)$
  the Bethe free-energy of the BP fixed point. Then
  \begin{align}
  \label{eq:Bethe-TN}
    e^{-\mcF_{Bethe}(\mcT)} 
      = \prod_{v}\Tr\big(T_v\prod_{u\in N_v}\hat{m}_{u\to v}\big) ,
  \end{align}
  where $\{\hat{m}_{u\to v}\}$ are a rescaling of the fixed-point
  messages, normalized such that 
  \begin{align}
  \label{eq:normalized-m}
    \Tr(\hat{m}_{u\to v}\cdot \hat{m}_{v\to u}) = 1 .
  \end{align}
\end{lemma}
The proof of \Lem{lem:Bethe} is given in Appendix~\ref{sec:Bethe}.
We note that the RHS of formula~\eqref{eq:Bethe-TN} is the product
of the contraction of small TNs, which in principle can be done
efficiently in parallel.

\subsection{The blockBP algorithm}

\begin{figure}
  \begin{center}
    \includegraphics[scale=1]{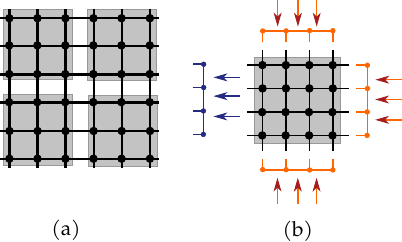}
  \end{center}
  \caption{Sketch of the blockBP algorithm for 2D tensor networks.
  (a) The 2D TN is partitioned into disjoint blocks of tensors.
  (b) The update rule for an outgoing message is the same as in
    regular BP, only that now the incoming messages are MPSs, and the contraction
    is done approximately using boundary MPS method, which results
    in an outgoing MPS message.} 
    \label{fig:blockBP-sketch} 
\end{figure}

The Bethe free-energy approximation of the BP algorithm to the
contraction of the TN is exact when the underlying graph is a
tree\cc{ref:MM2009-BPbook}. However, in the presence loops, its
quality might deteriorate, in particular when there are small loops
and strong correlations. Over the years, there have been many
attempts to improve this approximation by modifying/generalizing the
BP algorithm so that it would become more accurate on loopy
networks\cc{ref:YFW2000-GBP, ref:YFW2005-GBP, 
ref:Pelizzola2005-CVM, ref:Montanari2005-LCB, ref:Chertkov2006-LCB,
ref:Mooij2007-LCB, ref:LMRR2013-replica, ref:WZ2013-simpleGBP,
ref:ZW2015-LoopBP, ref:CantwellNewman2019-LoopyBP,
ref:KirkleyNewman2021-LoopyBP}. Usually, this is done at the expense
of higher computational costs. 

blockBP is one such algorithm, designed to be run on planar
tensor-networks. It relies on a simple idea, which is demonstrated
in \Fig{fig:blockBP-sketch}. We start by partitioning the TN into
disjoint blocks (\Fig{fig:blockBP-sketch}a) and think of the blocks
as the nodes of a new, ``coarse-grained'' TN. We then run the
regular BP on it by passing messages between neighboring blocks.
However, in order to avoid dealing with exponentially large messages
and block tensors, we represent a message $m_{\bar{v}\to \bar{u}}$
between block $\bar{v}$ and $\bar{u}$ as an \emph{MPS message} of
some maximal bond dimension $\chi$. Similarly, we do \emph{not} fuse
the tensors inside each block, and change the update rule in
\Eq{eq:BP-update} to a contraction of a small TN (made of the block
together with the incoming MPSs). This contraction is done with with
the boundary MPS method with maximal truncation bond $\chi$.
Finally, to estimate the Bethe free-energy approximation to the
global TN contraction, we use \Lem{lem:Bethe}, which remains
efficient also when using blocks and MPS messages.

The advantage of blockBP over the regular BP is that in blockBP the
short range loops within each block are treated more accurately (at
least up to truncation errors within the block contraction). The
larger we take the blocks, the more accurate the approximation
becomes at the expense of having to contract larger and larger
blocks. At the limit where the system becomes a single large block,
we recover the bMPS algorithm. As we shall see, for the purpose of
decoding the surface code, small blocks of size $6\times 6$ give
excellent results for $d\lessapprox 25$. Crucially, the blockBP
algorithm, just like the BP algorithm, can be run in parallel, and
this what makes it suitable for real-time decoding.

\subsection{Description of the blockBP decoder}
\label{sec:blockBP-decoder-desc}
We now describe the blockBP decoder for the surface code.

\noindent\textbf{Decoder Parameters:}
\begin{itemize}
  \item Code size $d$
  \item Local Pauli noise model
  \item BP threshold parameters: $\delta_0<\delta_1$ and
    \texttt{max-iter}.
  \item Local block size $k$
  \item Maximal MPS bond dimension $\chi$
\end{itemize}

\begin{algorithm}[H]

\SetKwInOut{Input}{input}\SetKwInOut{Output}{output}
\Input{A syndrome string $\us$ of size
$m=2d(d-1)$}
\Output{A guess for an operator $f_\us
  \bar{L}$ (where $\bar{L}$ in $\{\bar{I}, \bar{X}, \bar{Y},
  \bar{Z}\}$) with the maximal coset probability
  $\Tr(f_\us\bar{L}\mcG)$}
\BlankLine
{$f_\us\leftarrow $ a Pauli string that produces the syndrome $\us$}\;
\For{$\bar{L} \in \{\bar{I}, \bar{X}, \bar{Y}, \bar{Z}$\}}{
{$T_\us(\bar{L})\leftarrow$ Using the noise model and $f_\us$ as in \cRef{ref:BSMV2014-TN-decoding}}\;
{Partition $T_\us(\bar{L})$ into $k\times k$ blocks}\;
{Initialize MPS messages}\;
\For(\tcp*[h]{BP loop}){$t\leftarrow 1$ \KwTo $ \texttt{max-iter}$}{
{calculate the outgoing MPS message accordind to \Eq{eq:BP-update} using bMPS with $\chi$}\;
{Normalize outgoing messages}\;
\If {$\Delta<\delta_0$}{ 
{exit the BP loop}\;
}
}
\If {$\Delta<\delta_1$}{ 
{calculate the approximated $\Tr(T_\us(\bar{L}))$ via \Lem{lem:Bethe}}\;
}
}
\eIf {$\exists\; \bar{L} $ with $ \Delta<\delta_1$}{ 
{return $f_\us\bar{L}$ with $\Tr(T_\us(\bar{L}))_{max}$}\;
}
{
{return $f_\us\bar{L}$ with $\Delta_{min}$}\;
}

\caption{blockBP decoder}\label{algo_disjdecomp}
\end{algorithm}
\BlankLine
\noindent\textbf{Remarks:}
\begin{enumerate}
    \item  If $k$ does
        not divide $2d-1$ cleanly, the blocks at the edges of the TN
        can be of smaller sizes.
    \item We initialize all block MPS messages to 
      represent a uniform probability distribution.
      
    \item We normalize each outgoing message by making it 
      left-canonical with unit $L_2$ norm.
      
    \item We define the average distance between 
      the new messages and the previous messages: if $M$ is the
      total number of messages, the average distance is
      given by
      \begin{align}
        \Delta \EqDef \frac{1}{M}\Big(
          \sum_{i=1}^M \big\| \ketbra{m^{(t)}_i}{m^{(t)}_i} 
            -\ketbra{m^{(t-1)}_i}{m^{(t-1)}_i}\,\big\|^2\Big)^{1/2}
      \end{align}
  \item The running time complexity of a serial BP algorithm over $N$ 
    nodes with \texttt{max-iter} iterations and $T$ steps to
    calculate the outgoing messages of each node is
    $O(N\cdot\texttt{max-iter}\cdot T)$, and about
    $O(\texttt{max-iter}\cdot T)$ if we run all nodes in parallel.
    In our case, for a given code distance $d$, a block size $k$ and
    maximal bond dimension $\chi$, we have $N=O(d^2/k^2)$ and
    $T=O(k^2\chi^3)$. This gives us a serial running time of
    $O(n\chi^3\cdot\texttt{max-iter})$ --- which is identical or
    higher than the regular TN decoder. On the other hand,
    neglecting messages overhead, the parallel running time of the
    decoder should be of the order $O(\texttt{max-iter}\cdot
    k^2\chi^3 + n)$ --- which is potentially much faster. The
    additive $n$ term in the running time stems from unavoidable
    pre- and post- processing. We note that the running time
    can be further reduced by a factor of $4$ if we calculate the
    outgoing messages of each block in parallel.
    
    Finally, we remark that for near-future devices, these
    asymptotic estimates seem less important; the running time would
    probably be more influenced by the overall factors hidden in the
    big-O notation. These details are highly dependent on the
    particular implementation and underlying hardware.

  \item The algorithm uses two convergence error thresholds
    $\delta_0<\delta_1$. After running the BP loop for the coset
    $\bar{L}f_s\mcG$, if the final messages have a convergence error
    $\Delta<\delta_1$, then we consider it sufficiently converged so
    that we can trust its Bethe free-energy approximation for
    $\Tr(\bar{L}f_s\mcG)$. However, if non of the cosets produced
    $\Delta<\delta_1$, we pick the one with the smallest $\Delta$.
    This is because, as we noted empirically, the right coset is
    most often the one with the quickest convergence.
    
  \item In order to facilitate the convergence we use \emph{damping}
    --- a common practice in BP algorithms. Given a damping
    parameter $\eta$, if $\ket{m^{(t+1)}}$ is the new outgoing MPS
    message and $\ket{m^{(t)}}$ is the old outgoing MPS message, we
    replace $\ket{m^{(t+1)}}$ by
    \begin{align}
      \ket{m^{(t+1)}}_{damp} \EqDef (1-\eta)\ket{m^{(t+1)}} 
        + \eta\ket{m^{(t)}} .
    \end{align}
    Throughout our simulations, we used $\eta=0.1$
    
  \item Another important BP aspect is \emph{scheduling} ---
    determining the order at which the different modes will send and
    receive messages. Here we use a simple ``two-layers'' approach.
    We color the blocks in black and white as a chessboard, and at
    each round, either the black blocks send messages to the white
    blocks or the other way around. Withing a given round, all
    updates can be done in parallel.
    
  \item For small $k$, it might be computationally advantageous to 
    run a ``nav\"ie blockBP algorithm'', in which we simply fuse the
    vertices of the blocks to one mega-vertex, and run plain BP
    between these vertices. This way, we work with tensors of much
    higher bond dimension, but we save the overhead of the MPS
    machinery, and eliminate the MPS truncation error that is
    associated with finite $\chi$. Whether or not this is beneficial
    depends on implementation and the hardware. In this work, we
    used this approach for the $k=1,2,4$ cases. For the $k=6$ case,
    we first passed to a lattice that was the result of fusing 
    $3\times 3$ vertices of the original lattice, and used blockBP
    with $k'=2$. See \Sec{sec:numerics} for more details.

\end{enumerate}

\section{Numerical simulations}
\label{sec:numerics}

To test the performance of the blockBP decoder, we have performed
several Monte Carlo experiments in the code capacity model, in which
flawless stabilizer measurements are assumed.  We leave the
important question of how this decoder performs in the presence of
noisy stabilizer measurements (i.e., in the circuit noise model) for
future research. Our raw numerical data and source code can provided
upon reasonable request.

Our benchmark included lattice sizes $d=5,9,13,17,21,15$ and
different levels of single-qubit depolarizing noise
\begin{align*}
  N(\rho) = (1-\eps)\rho + \frac{\eps}{3}\big(X\rho X + Y\rho Y
  +Z\rho Z)
\end{align*}
for $\eps=0.08, 0.09, \ldots, 0.15$. In all runs we used the
following parameters:
\begin{itemize}
  \item \textbf{Block size:} $k=1,2,4,6$
  
  \item \textbf{Convergence parameters:} 
    \begin{itemize}
      \item \texttt{max-iter}=20
      \item $\delta_0=10^{-4}$, $\delta_1=10^{-2}$
    \end{itemize}
    
  \item \textbf{Damping:} $\eta=0.1$
  
  \item \textbf{MPS truncation:} $\chi=16$
\end{itemize}
As mentioned at the end of \Sec{sec:blockBP-decoder-desc}, to speed
up the decoder we used a plain BP algorithm for the $k=1,2,4$ cases
by fusing a $k\times k$ block into a single tensor with $4$ legs of
dimension $2^k$. For the $k=6$ case, we first moved to a system in
which $3\times 3$ blocks were contracted to a single tensor with a
bond dimension of $D'=2^3=8$, and then ran blockBP with $k'=2$ and
$\chi=16$.

For reference, we have compared the performance of the blockBP
decoder to that of the MWPM decoder and a bMPS based decoder with
$\chi=16$. All runs were performed using the \texttt{qecsim} Python
library\cc{ref:qecsim2020}. For the blockBP simulation, we used our
own Python code, which was incorporated into \texttt{qecsim} as a
decoder. For every set of parameters (decoder, $d$, $k$, $\eps$),
which produced an empirical logical error probability $\hat{P}_L$,
we have used $N\ge \min(30,000, \ P_L/100)$ shots to estimate $P_L$,
which amounts for at least $95\%$ confidence interval of
$[0.8\hat{P}_L, 1.25\hat{P}_L]$\cc{ref:Jeruchim1984-MCsamp}.

\begin{figure}
  \begin{center}
    \includegraphics[scale=0.75]{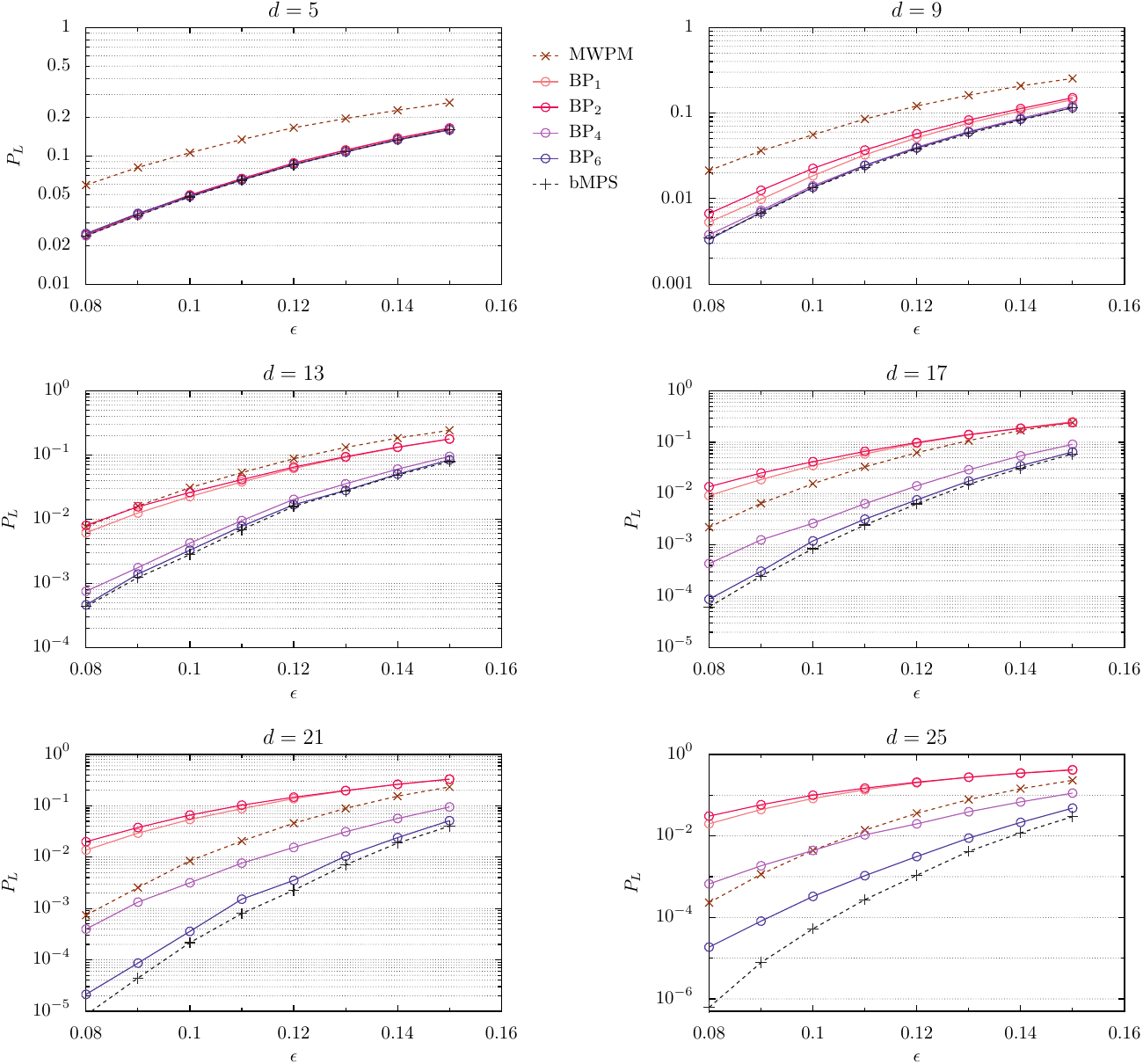}
  \end{center}
  \caption{The dependence of the logical probability error $P_L$ on
  the depolarization noise strength for $d=5,9, \ldots, 25$, as
  given by the blockBP, bMPS and MWPM decoders.} \label{fig:PL-d}
\end{figure}

\begin{figure}
  \begin{center}
    \includegraphics[scale=0.75]{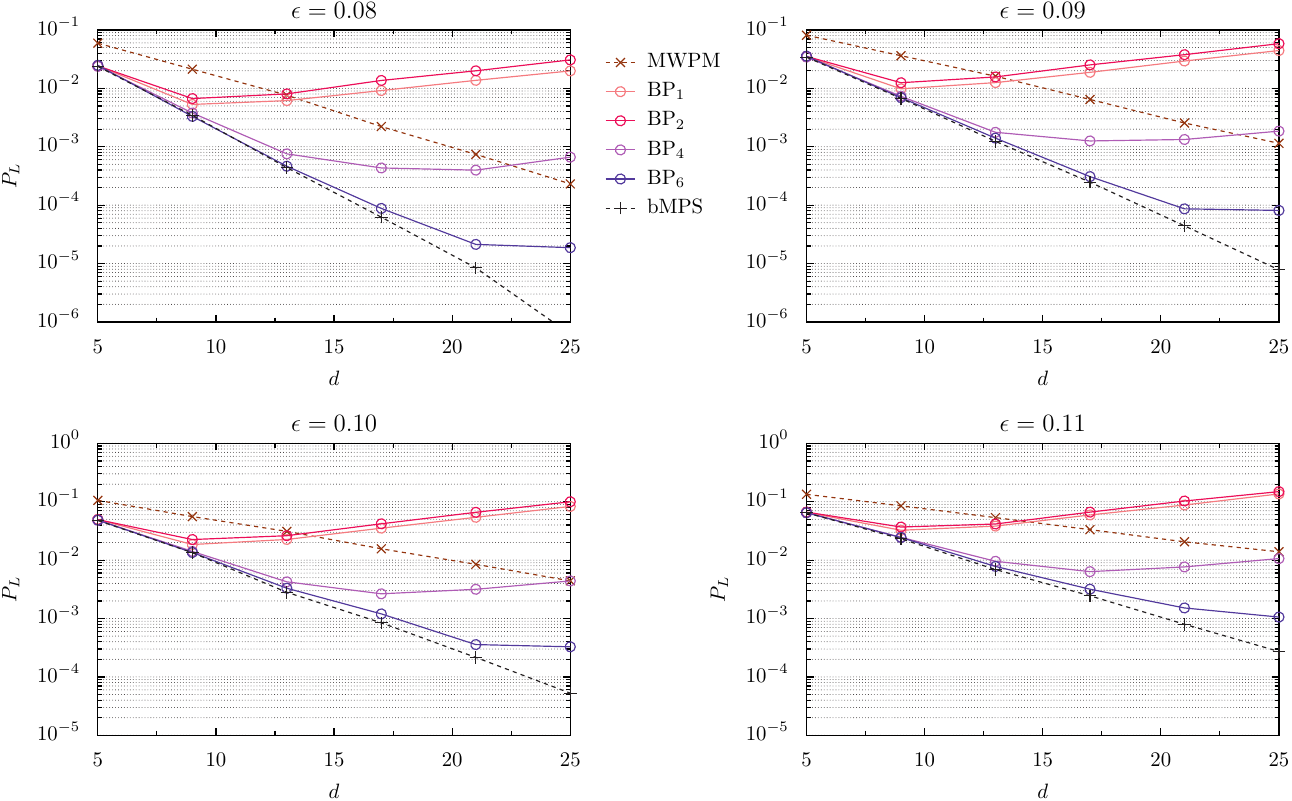}
  \end{center}
  \caption{The logical error probability $P_L$ as a function of the
  code distance $d$ for a physical depolarization rates of
  $\eps=0.08, 0.09, 0.10, 0.11$ and for the different decoders. }
  \label{fig:PL-I}
\end{figure}

\begin{figure}
  \begin{center}
    \includegraphics[scale=0.75]{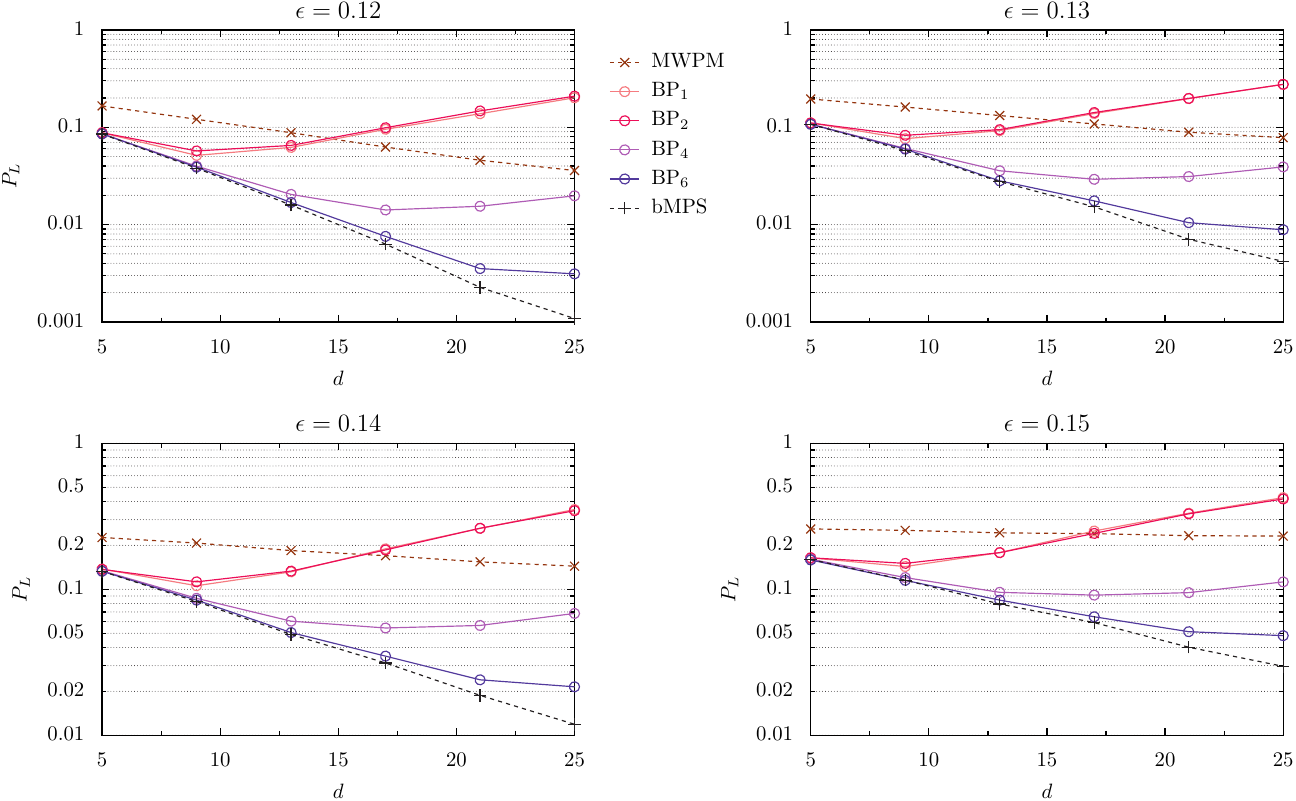}
  \end{center}
  \caption{The logical error probability $P_L$ as a function of the
  code distance $d$ for a physical depolarization rates of
  $\eps=0.12, 0.13, 0.14, 0.15$ and for different decoders. Every
  point was calculated using at least $100$ failures.}
  \label{fig:PL-II}
\end{figure}

In \Fig{fig:PL-d}, we plot the logical probability error $P_L$ as a
function of the depolarization noise strength $\eps$ for various
code sizes $d=5, 9, 13, \ldots, 25$. Perhaps the most evident
feature of the blockBP decoder is its sensitivity to the parameter
$d$. For $d=5$, the results of all blockBP runs are virtually
indistinguishable from the MPS decoder (which is nearly optimal).
However, as $d$ increases, the blockBP instances with smaller $k$
become less and less accurate. This phenomena is best captured in
Figs.~\ref{fig:PL-I},\ref{fig:PL-II}, where we plot the logical
failure probability as a function of $d$ for different noise levels.
We see that for every block size $k$ there exists a lattice size
$d_k$ at which $P_L$ stops decreasing as a function of $k$. This
minima is roughly independent of $\eps$. For $k=1,2$, we have
$d_k\approx 9$, whereas $d_4\approx 17$ and $d_6\approx 25$. As
\Fig{fig:PL-d} shows, when $d\lessapprox d_k$, the blockBP decoder
often yields logical error probabilities that are order of magnitude
smaller than those of the MWPM decoder. 

The degradation of decoding quality as a function of $d$ is not
unique to blockBP. It also exists in the usual TN decoder, which is
based on bMPS contraction, though to a much lesser degree. Indeed,
also in the TN decoder, if we let $d$ grow while keeping the
truncation bond $\chi$ fixed, the truncation errors in the bMPS
contraction will eventually dominate the estimated probability and
increase the failure rate. While we believe that this global
behavior is unavoidable, it might be possible to improve the decoder
at these regimes not only by increasing $k$, but also by increasing
the \texttt{max-iter} parameter, decreasing $\delta_0, \delta_1$
parameters, or using smarter BP scheduling\cc{ref:MJG2019-BP-sched}.
We leave these possibilities for future research.

We conclude this section by describing two empirical observations, 
which might be relevant for further improvements of the decoder. 

\paragraph{Fast convergence to the right coset.} 
Generally, when running the blockBP decoder, we noticed that the
right coset is almost always the one with the fastest BP
convergence. This is an indication that the TN that represents the
right coset is less critical, with fewer long range correlations. In
our blockBP decoder we take advantage of this behavior by using the
$\delta_1$ parameter in conjunction with a relatively low
\texttt{max-iter} bound. Together, these parameters tend to
automatically reject cosets with a slow BP convergence --- which are
mostly the wrong cosets.  This results in many cases where the
blockBP is seemingly performing much better than one would have
guessed by comparing the coset probabilities calculated by blockBP
to the more accurate ones calculated by bMPS. This suggests that
there is more structure in the TN that can be used to infer the
right coset, than merely its overall contraction value. Finally, we
remark that a similar behavior was also observed in the bMPS decoder
in \cRef{ref:BSMV2014-TN-decoding}, where the convergence of the
bMPS algorithm with respect to the truncation bond $\chi$ is much
faster on the right coset than on the other cosets. 
\paragraph{Faster convergence for $\eps\to 0$.}
As $\eps\to 0$, the TNs of the cosets become less critical, with
shorter range correlations, and as a result the BP messages converge
faster and produce more accurate results. This means that for low
noise rate we can lower the \texttt{max-iter} parameter without
substantially sacrificing the decoder performance.

\newpage

\onecolumngrid

\twocolumngrid

{~}

\newpage

\section{Summary and conclusions}
\label{sec:conclusions}

In this paper we presented a new decoder to the surface code and
other topological 2D codes. It is based on the framework of
tensor-network decoding, equipped with the parallelism of the
belief-propagation algorithm in the form the \blockBP algorithm.
Unlike traditional uses of BP in the decoding problem, our BP
algorithm does not run on the underlying Tanner graph; instead it is
used to approximate the contraction of the tensor networks of the
error cosets, which makes it a DQMLD decoder. The central parameter
in the decoder is the block size $k$, which determines both the
complexity of the decoding and its accuracy. In our Monte-Carlo
simulations, using only $20$ message passing rounds, we have shown
that for code distances of $d\lessapprox 9$, a block size of $k=1,2$
exceeds the accuracy of the MWPM decoder. Similarly, for
$d\lessapprox 17$ with $k=4$ and $d\lessapprox 25$ with $k=6$. Being
a BP algorithm, our decoder uses a small number of message-passing
rounds to converge, which potentially makes it suitable for
real-time decoding.

Our research leaves several open questions. First, it is crucial to
see if our decoder can be extended to the circuit noise model, and
test its performance in that setting. A recent
paper\cc{ref:CCR2023-TNdec} showed that the tensor-network decoding
framework can be adapted to this setting, which leads to believe
that so is the \blockBP decoder. An equally important question is to
understand if the decoder can actually be implemented in real-time
decoding. Since belief-propagation algorithms have long been used in
such scenarios, we believe that, at least in the noiseless
measurement model, this should be possible. It would also be
interesting to see the performance of our decoder at different noise
models, for example, biased noise. Finally, there are many different
techniques in the literature to improve the performance of BP
algorithms, such as smarter scheduling\cc{ref:MJG2019-BP-sched}, or
using techniques of machine learning\cc{ref:NBB2016-neuralBP}. It
would be interesting to if any of these methods can improve the
performance of our decoder.


\begin{acknowledgments}
  We thank the computational support of D.~Poletti, B.~Xing, and
  X.~Xu. from SUTD, Singapore, and the support from the joint
  Israel-Singapore NRF-ISF Research grant NRF2020-NRF-ISF004-3528.
\end{acknowledgments}


\appendix

\section{Proof of \Lem{lem:Bethe}}
\label{sec:Bethe}

We begin by introducing the Bethe free-energy formula in the TN
notation. Recall that with every edge $e=(u,v)\in E$, we associated
a variable $x_e$ that takes values in the range of the tensor
indices that are contracted along that edge. Then the message
$m_{u\to v}(x_e)$ can be viewed either as a vector or as a function
of $x_e$. Similarly, every tensor $T_v$ with adjacent edges $e_1,
e_2, \ldots$ can be viewed as a function $T_v(x_{e_1}, x_{e_2},
\ldots) \EqDef T_v(\Bx_v)$, where we have defined $\Bx_v \EqDef
(x_{e_1}, x_{e_2}, \ldots )$ to be the set of edge variables that
are associated with the tensor $T_v$. Using this notation, we define
the \emph{edge marginals} $\{P_e(x_e)\}$ and the vertex marginals
$\{P_v(\Bx_v)\}$ by
\begin{align}
  P_e(x_e) &\EqDef \frac{1}{Z_e} m_{u\to v}(x_e) 
    \cdot m_{v\to u}(x_e), \\
  P_v(\Bx_v) &\EqDef \frac{1}{Z_v} T_v(\Bx_v)
    \cdot \prod_{u\in N_v} m_{u\to v}(x_e), 
\end{align}
where $Z_e, Z_v$ is a normalization factors defined such that
$\sum_{x_e} P_e(x_e) = \sum_{\Bx_v}P(\Bx_v)=1$. We note that when
the TN has negative or complex values, the ``marginals'' above need
not be actual probability distributions. Our formula will
nevertheless hold also in these cases. Note also that the edge
marginal $P_e(x_e)$ can also be expressed using the normalized
messages as 
\begin{align}
\label{eq:e-marginal}
  P_e(x_e) = \hat{m}_{u\to v}(x_e) \cdot \hat{m}_{v\to u}(x_e).
\end{align}

As shown in \cRef{ref:AA2021-TNBP} (Eq.~A12), in terms of $P_e(x_e)$
and $P_v(\Bx_v)$, the Bethe free-energy of the TN is given by
\begin{align}
\label{eq:Bethe-orig}
  F_{Bethe}(\mcT) &= \sum_{v\in V} \sum_{\Bx_v}
    P_v(\Bx_v)\ln \frac{P_v(\Bx_v)}{T_v(\Bx_v)}\\
   &- \sum_{e \in E} \sum_{x_e} P_e(x_e)\ln P_e(x_e). \nonumber
\end{align}
It can be used to approximate  contraction of $\mcT$ by the 
formula $\Tr(\mcT) \simeq e^{-F_{Bethe}}(\mcT)$, which becomes exact
when the underlying TN has a tree structure\cc{ref:MM2009-BPbook}.

\begin{figure}
  \begin{center}
    \includegraphics[scale=1]{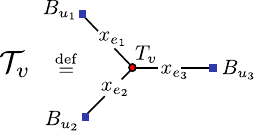}
  \end{center}
  \caption{The auxiliary tensor network $\mcT_v$ that is constructed
  for every vertex $v\in V$. The tensor sitting at $v$ is the
  tensor $T_v$ form the original TN. The leaf tensors $B_{u_2},
  B_{u_2}, B_{u_3}$ are defined by the incoming messages to $v$:
  $B_{u_i}(e_i) \EqDef m_{u_i\to v}(x_{e_i})$.}
  \label{fig:auxiliary-graph}
\end{figure}

For each tensor $T_v$, we now construct a star-like tensor-network
$\mcT_v$ in which $T_v$ sits at the center, surrounded by its
neighboring vertices from $G$. On each neighboring vertex $u\in
N_v$, we place a one-leg tensor $B_u$ that is defined using the
incoming \emph{normalized} messages $\hat{m}_{u\to v}(x_e)$ (see
condition \eqref{eq:normalized-m}):
\begin{align*}
  B_u(x_e) \EqDef \hat{m}_{u\to v}(x_e) .
\end{align*}
This construction is depicted in \Fig{fig:auxiliary-graph}.

We now observe:
\begin{enumerate}
  \item The contraction of $\mcT_v$ is:
    \begin{align*}
      \Tr(\mcT_v) &= \Tr(T_v \cdot B_{u_1}\cdot B_{u_2} \cdots)
      = \Tr\big(T_v \prod_{u\in N_v}\hat{m}_{u\to v}\big) .
    \end{align*}
    
  \item As $\mcT_v$ is a tree TN, its contraction can be written 
    exactly using its Bethe free-energy,
    \begin{align*}
      \Tr(\mcT_v) = e^{-F_{Bethe}(\mcT_v)} .
    \end{align*}
\end{enumerate}
Using these observations it follows that to prove the
lemma it is sufficient to show that 
\begin{align}
\label{eq:decompose-Bethe}
  \mcF_{Bethe}(\mcT) = \sum_v \mcF_{Bethe}(\mcT_v),
\end{align}
as it would imply
\begin{align*}
  e^{-\mcF_{Bethe}(\mcT)} &= \prod_v e^{-\mcF_{Bethe}(T_v)} \\
   &= \prod_v\Tr\big(T_v \prod_{u\in N_v}\hat{m}_{u\to v}\big) .
\end{align*}
The rest of the proof is therefore dedicated to proving
\Eq{eq:decompose-Bethe}.

We begin by  writing
$F_{Bethe}(\mcT_v)$ as a sum of 3 terms $A,B,C$:
\begin{align}
\label{eq:Bethe-Tv}
  &F_{Bethe}(\mcT_v) = \overbrace{\sum_{\Bx_v}
    P_v(\Bx_v)\ln
      \frac{P_v(\Bx_v)}{T_v(\Bx_v)}}^{A} \\
   &- \overbrace{\Big(\sum_{x_{e_1}} P_{e_1}(x_{e_1})
     \ln P_{e_1}(x_{e_1}) 
     + \sum_{x_{e_2}} P_j(x_{e_2})\ln P_i(x_{e_2}) 
       + \ldots\Big)}^B \nonumber \\
   &+\overbrace{\sum_{x_{e_1}} P_{u_1}(x_{e_1}) 
     \ln \frac{P_{u_1}(x_{e_1})}{B_{u_1}(x_{e_1})}
   + \sum_{x_{e_2}} P_{u_2}(x_{e_2}) 
     \ln \frac{P_{u_2}(x_{e_2})}{B_{u_2}(x_{e_2})} + \ldots}^C.
     \nonumber
\end{align}

Importantly, up to normalization, the converged BP messages on
$\mcT_v$, coincide with those of the original TN. Therefore, the
vertex edge marginal $P_v(\Bx_v)$ of $\mcT_v$ coincides with that of
the original $\mcT$, and the same thing is true for the edge
marginals $P_{e_i}(x_{e_i})$. The vertex marginals
$P_{u_i}(x_{e_i})$, however, which appear in $C$, are not identical
to those in the original TN. In fact, they may not even be defined
on the same number of variables. Nevertheless, we note that by 
\Eq{eq:e-marginal},
\begin{align*}
  \frac{P_{u_i}(x_{e_i})}{B_{u_i}(x_{e_i})} =
	\frac{\hat{m}_{u_i\to v}(x_{e_i})
      \cdot \hat{m}_{v\to u_i}(x_{e_i})}{\hat{m}_{u_i\to v}(x_{e_i})}
	= \hat{m}_{v\to u_i}(x_{e_i})
\end{align*}
and therefore $C$ becomes
\begin{align*}
  &\sum_{x_{e_1}}P_i(x_{e_1})\ln\hat{m}_{v\to u_1}(x_{e_1})
	  + \sum_{x_{e_2}}P_j(x_{e_2})\ln\hat{m}_{v\to u_2}(x_{e_2}) \\
        & \ + \ldots
\end{align*}  
Summing $F_{Bethe}(\mcT_v)$ contributions over all nodes $v$, we see
that term $A$ gives us exactly the corresponding expression in
$F_{Bethe}(\mcT)$ as in \Eq{eq:Bethe-orig}. Term $B$ gives us
$-2\sum_{e\in E} P_e(x_e)\ln P_e(x_e)$ because for each edge $e$
there is a contribution from each of its two adjacent nodes.
However, for each edge $e=(u,v)$, the term $C$ contains
\begin{align*}
  &\sum_{x_e} P_e(x_e)\big(\ln \hat{m}_{u\to v}(x_e)
      + \ln \hat{m}_{v\to u}(x_e)\big) \\
  & \ = \sum_{x_e} P_e(x_e)\ln P_e(x_e) ,
\end{align*}
where the last equality follows from \Eq{eq:e-marginal}.
Consequently, the extra term in $B$ in canceled and this concludes
the proof.

\bibliographystyle{ieeetr}
\bibliography{BLOCKD}

\end{document}